\PassOptionsToPackage{unicode}{hyperref}
\PassOptionsToPackage{hyphens}{url}
\documentclass[
  11pt,
]{svjour3}
\usepackage{lmodern}
\usepackage{amssymb,amsmath}
\usepackage{ifxetex,ifluatex}
\ifnum 0\ifxetex 1\fi\ifluatex 1\fi=0 
  \usepackage[T1]{fontenc}
  \usepackage[utf8]{inputenc}
  \usepackage{textcomp} 
\else 
  \usepackage{unicode-math}
  \defaultfontfeatures{Scale=MatchLowercase}
  \defaultfontfeatures[\rmfamily]{Ligatures=TeX,Scale=1}
\fi
\IfFileExists{upquote.sty}{\usepackage{upquote}}{}
\IfFileExists{microtype.sty}{
  \usepackage[]{microtype}
  \UseMicrotypeSet[protrusion]{basicmath} 
}{}
\makeatletter
\@ifundefined{KOMAClassName}{
  \IfFileExists{parskip.sty}{%
    \usepackage{parskip}
  }{
    \setlength{\parindent}{0pt}
    \setlength{\parskip}{6pt plus 2pt minus 1pt}}
}{
  \KOMAoptions{parskip=half}}
\makeatother
\usepackage{xcolor}
\IfFileExists{xurl.sty}{\usepackage{xurl}}{} 
\IfFileExists{bookmark.sty}{\usepackage{bookmark}}{\usepackage{hyperref}}
\hypersetup{
  pdftitle={Having a Ball: evaluating scoring streaks and game excitement using in-match trend estimation},
  hidelinks,
  pdfcreator={LaTeX via pandoc}}
\usepackage{graphicx,grffile}
\makeatletter
\def\maxwidth{\ifdim\Gin@nat@width>\linewidth\linewidth\else\Gin@nat@width\fi}
\def\maxheight{\ifdim\Gin@nat@height>\textheight\textheight\else\Gin@nat@height\fi}
\makeatother
\setkeys{Gin}{width=\maxwidth,height=\maxheight,keepaspectratio}
\makeatletter
\def\fps@figure{htbp}
\makeatother
\usepackage{bm}
\usepackage{amssymb}
\usepackage[labelfont=bf]{caption}

\usepackage{multirow}
\usepackage{float}
\floatstyle{plaintop}
\restylefloat{table}
\usepackage[thmmarks]{ntheorem}
\theoremstyle{nonumberplain}
\usepackage{xcolor}

\usepackage{fontawesome5}
\newcommand{\BTheta}{\ifthenelse{\value{bbsym4theta}=0}{\text{\faBasketballBall}}{\mathbf{\Theta}}}
\authorrunning{CT Ekstrøm and AK Jensen}
\titlerunning{Having a Ball}
\institute{Biostatistics, Department of Public Health\\University of Copenhagen\\ Øster Farimagsgade 5\\ 1014 København K\\ DENMARK}
\usepackage{booktabs}
\usepackage{longtable}
\usepackage{array}
\usepackage{multirow}
\usepackage{wrapfig}
\usepackage{float}
\usepackage{colortbl}
\usepackage{pdflscape}
\usepackage{tabu}
\usepackage{threeparttable}
\usepackage{threeparttablex}
\usepackage[normalem]{ulem}
\usepackage{makecell}

\title{Having a Ball: evaluating scoring streaks and game excitement using
in-match trend estimation}
\author{Claus Thorn Ekstrøm and Andreas Kryger Jensen\\
(0000-0003-1191-373X), (0000-0002-8233-9176)\\
Biostatistics, Department of Public Health, University of Copenhagen\\
\href{mailto:ekstrom@sund.ku.dk}{\nolinkurl{ekstrom@sund.ku.dk}},
\href{mailto:aeje@sund.ku.dk}{\nolinkurl{aeje@sund.ku.dk}}}
\date{22 December, 2020}

\begin{document}
\maketitle

\newpage

\newcounter{bbsym4theta}
\setcounter{bbsym4theta}{0}

\begin{abstract}

Many popular sports involve matches between two teams or players where
each team have the possibility of scoring points throughout the match. While
the overall match winner and result is interesting, it conveys little
information about the underlying scoring trends throughout the
match. Modeling approaches that accommodate a finer granularity of the
score difference throughout the match is needed to evaluate
in-game strategies, discuss scoring streaks, teams strengths, and other
aspects of the game.

We propose a latent Gaussian process to model the score difference
between two teams and introduce the Trend Direction Index as an easily
interpretable probabilistic measure of the current trend in the match
as well as a measure of post-game trend evaluation. In addition we
propose the Excitement Trend Index --- the expected number of
monotonicity changes in the running score difference ---
as a measure of overall game excitement.

Our proposed methodology is applied to all 1143 matches from
the 2019--2020 National Basketball Association (NBA) season. We show
how the trends can be interpreted in individual games and how the
excitement score can be used to cluster teams according to how
exciting they are to watch.
\end{abstract}

\begin{center}
\end{center}

\hypertarget{declarations}{%
\section*{Declarations}\label{declarations}}
\addcontentsline{toc}{section}{Declarations}

\textbf{Funding}: The research was funded by the University of
Copenhagen.

\textbf{Conflicts of interest/Competing interests}: None

\textbf{Availability of data and material}: All data are available at
\url{https://github.com/aejensen/Having-a-Ball}

\textbf{Code availability}: All code are available at
\url{https://github.com/aejensen/Having-a-Ball} and
\url{https://github.com/aejensen/TrendinessOfTrends}.

\newpage

\hypertarget{introduction}{%
\section{Introduction}\label{introduction}}

Sports analytics receive increasing attention in statistics and not just
for match prediction or betting but also for game evaluation, in-game
and post-game coaching purposes, and for setting strategies and tactics
in future matches.

Many popular sports such as football (soccer), basketball, boxing, table
tennis, volleyball, American football, and handball involve matches
between two teams or players where each team have the possibility of
scoring points throughout the match. Several research papers seek to
predict the end match result (e.g., Karlis and Ntzoufras (2003); Groll
et al. (2019); Gu and Saaty (2019); Cattelan, Varin, and Firth (2013))
in order to infer the match winner and potentially the winner of a
tournament (Ekstrøm et al. (2020); Baboota and Kaur (2018)). While the
overall match result is highly interesting it conveys very little
information about the individual development and trends throughout the
match and modeling approaches that allow finer granularity of the
running score difference throughout the match are needed.

The trend in the score difference between the two teams is a proxy for
their underlying strengths. In particular, sustained periods of time
where the score difference increases suggest that one team outperforms
the other whereas periods where the teams are constantly catching up to
each other suggest that the teams' strengths in those periods are
similar. Modeling the local trend of the score difference will therefore
reflect several aspects of the game, in particular, the team strengths
and game dynamics and momentum as they develop through the match.

\begin{figure}[htb]
\center
\includegraphics{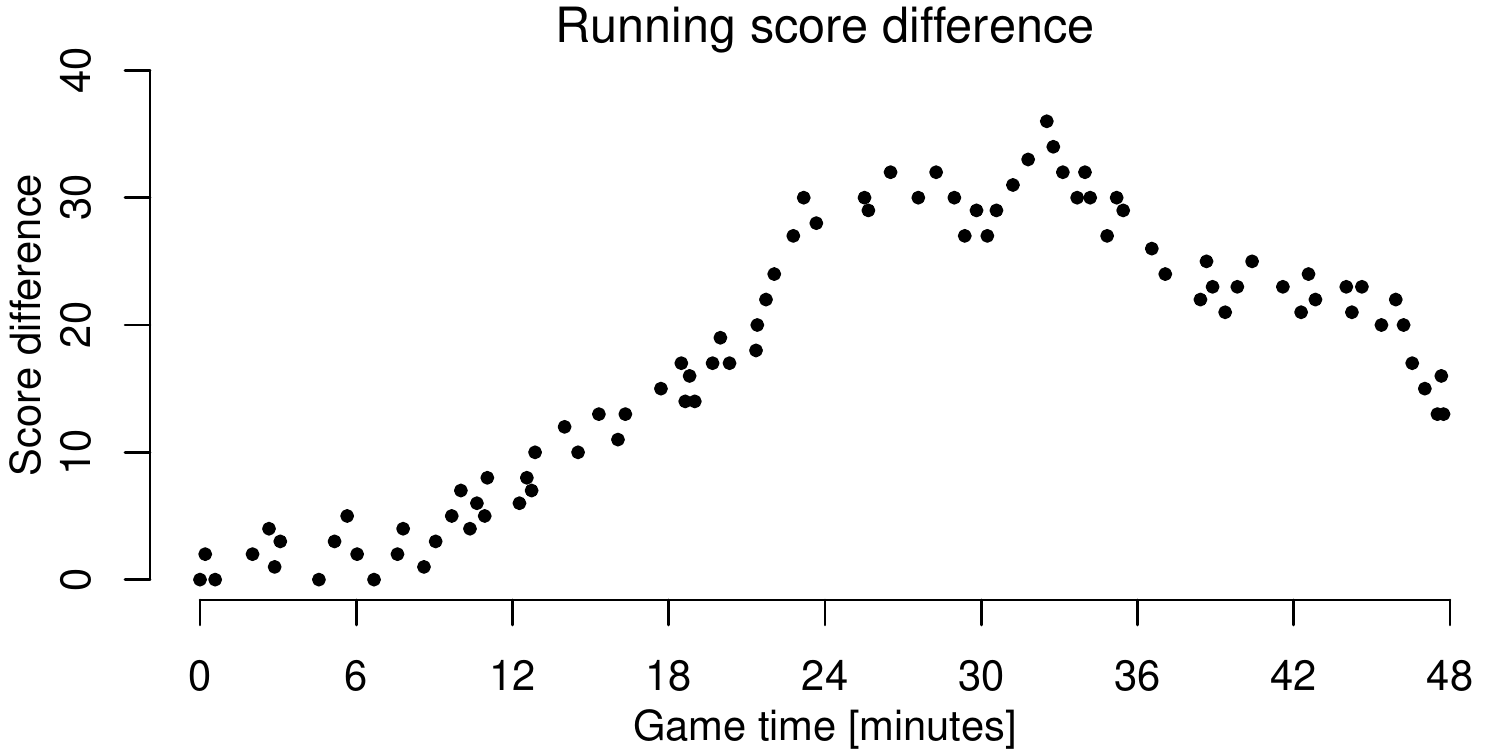}
\caption{Game development in the final match of the NBA 2019--2020 season between Los Angeles Lakers and Miami Heat at October 11, 2020. Positive values indicate that LA Lakers are leading.}
\label{fig1}
\end{figure}

Figure \ref{fig1} shows the running score difference for the final match
of the playoffs in the 2019--2020 National Basketball Association (NBA)
series between Los Angeles Lakers and Miami Heat. Positive numbers
indicate that LA Lakers are leading and the running score difference
shows that the Lakers pulled ahead until the third quarter where Miami
Heat started to keep up the scoring pace before overtaking the Lakers
and reducing the lead.

In this paper, we will consider the score difference between two teams
as a latent Gaussian process and use the Trend Direction Index (TDI)
from Jensen and Ekstrøm (2020b) as a measure to evaluate the local
probability of the \emph{monotonicity} of the latent process at a given
time point during a match. The Trend Direction Index uses a Bayesian
framework to provide a direct answer to questions such as ``What is the
probability that the latent process is increasing (i.e., that one team
is doing better than another) at a given time-point?''. This will allow
real-time evaluation of the score difference trend at the current
time-point in-game and will provide post-game inference about the
``hot'' periods of a match where one team out-performed the other.
Furthermore, we present the Excitement Trend Index (ETI) as an objective
measure of spectator excitement in a given match. The ETI is defined as
the expected number of times that the score difference changes
monotonicity during a match. If the score difference changes
monotonicity often then that echos a game where both teams frequently
score whereas a game with a low ETI will represent a one-sided match
where one team is doing consistently better than the other over
sustained periods of time.

Other authors have considered using continuous processes to model the
score difference of matches. Gabel and Redner (2012) shows that NBA
basketball score differences are well described by a continuous-time
anti-persistent random walk which suggests that a latent Gaussian
process might be viable. Chen, Dawson, and Müller (2020) consider a
functional data model for dynamic behavior of cross-sectional ranks over
time. While this approach can disentangle the individual and population
effect on the ranks of the individual teams over time its setup is not
really geared towards analyzing single matches. We use an idea similar
to Chen and Fan (2018) but they do not have the same underlying Gaussian
intensity process that enables us to make various Bayesian probabilistic
statements throughout and after the game.

The paper is structured as follows. In the next section we introduce
trend modeling of score differences through a latent Gaussian process
and define the Trend Direction Index and the Excitement Trend Index that
capture the local trends in monotonicity and game excitement,
respectively. In Section \ref{sec:application} we apply our proposed
methodology to analyze both the final match of the playoff as well as
evaluating the game excitement distribution of the season by considering
the ETIs from all 1143 matches from the 2019--2020 NBA season. We show
how this distribution can be used to assess relative match excitement
and how the ETI can be used to classify teams according to their average
level of match excitement. We conclude with a discussion in Section
\ref{sec:discussion}. Materials to reproduce this manuscript and its
analyses can be found at Jensen and Ekstrøm (2020a).

\hypertarget{sec:method}{%
\section{Methodology}\label{sec:method}}

Our model is based on the observed score differences \(D_m(t)\) in a
given match indexed by \(m\) and time \(t\). For each match we observe
the random variables
\(\mathcal{D}_m = \left(t_{mi}, D_{mi}\right)_{0 < i \leq J_m}\) where
\(t_{m1} < t_{m2} < t_{mi} < \ldots < t_{m J_m}\) are the ordered time
points at which any team scores, \(D_{mi} = D_m(t_{mi})\) is the
associated difference in scores at time \(t_{mi}\), and \(J_m\) is the
total number of scorings during the match. We use the convention that
\(D_m\) is the difference in scores of the away team with respect to the
home team so that \(D_m(t) > 0\) means that the away team is leading at
time \(t\).

We assume that the observed data from a given match are noisy
realizations of a latent smooth, random function defined in continuous
time and evaluated at the random time points where scorings occur. Let
\(d_m\) be the latent function from which the realizations
\(\mathcal{D}_m\) are generated. Our objective is to infer \(d_m\) and
its time dynamics from \(\mathcal{D}_m\). In pursuance of this ambition
we propose the following model where \(d_m\) is a Gaussian process
defined on a compact subset of the real line \(\mathcal{I}_m\)
corresponding to the duration of the \(m\)'th game, and the observed
data conditional on the scoring times and the values of the latent
process at these times are independently normally distributed random
variables with a match specific variance. This model can be stated
hierarchically as \begin{align}
\begin{split}
  \ifthenelse{\value{bbsym4theta}=0}{\text{\faBasketballBall}}{\mathbf{\Theta}}_m \mid \bm{\Psi}_m, \mathbf{t}_m &\sim H(\ifthenelse{\value{bbsym4theta}=0}{\text{\faBasketballBall}}{\mathbf{\Theta}}_m \mid \bm{\Psi}_m)\\
  d_m(t) \mid \ifthenelse{\value{bbsym4theta}=0}{\text{\faBasketballBall}}{\mathbf{\Theta}}_m &\sim \mathcal{GP}(\mu_{\bm{\beta}_m}(t), C_{\bm{\theta}_m}(s, t))\\
  D_m(t_{mi}) \mid d_m(t_{mi}), t_{mi}, \ifthenelse{\value{bbsym4theta}=0}{\text{\faBasketballBall}}{\mathbf{\Theta}}_m &\overset{iid}{\sim} N\left(d_m(t_{mi}), \sigma_{m}^2\right)
\end{split}
\label{eq:hierarchicalmodel}
\end{align} where
\(\ifthenelse{\value{bbsym4theta}=0}{\text{\faBasketballBall}}{\mathbf{\Theta}}_m = (\bm{\beta}_m, \bm{\theta}_m, \sigma_{m}^2)\)
is a vector of hyper-parameters governing the dynamics of the latent
Gaussian process with a prior distribution \(H\) indexed by parameters
\(\bm{\Psi}_m\), and \(\mathbf{t}_m = (t_{m1}, \ldots, t_{m J_m})\) is
the vector of time points where scorings occur in the match. The
functions \(\mu_{\bm{\beta}_m}\) on \(\mathcal{I}_m\) and
\(C_{\bm{\theta}_m}\) on \(\mathcal{I}_m \times \mathcal{I}_m\) are the
prior mean and covariance functions of the latent Gaussian process, and
\(\sigma_m^2\) is the variance characterizing the magnitude of the
deviations between for the observed score differences and the values of
the latent process.

A Gaussian process is characterized by the multivariate joint normality
of all of the joint distributions resulting from evaluating the process
at any finite set of time points (Rasmussen and Williams (2006)).
Specifically, for any finite set
\(\mathbf{t}^\ast \subset \mathcal{I}_m\) it follows that the vector
\(d_m(\mathbf{t}^\ast) \mid \ifthenelse{\value{bbsym4theta}=0}{\text{\faBasketballBall}}{\mathbf{\Theta}}_m\)
is distributed as
\(N(\mu_{\bm{\beta}_m}(\mathbf{t}^\ast), C_{\bm{\theta}_m}(\mathbf{t}^\ast, \mathbf{t}^\ast))\)
where \(\mu_{\bm{\beta}_m}(\mathbf{t}^\ast)\) is the vector generated by
evaluating the prior mean function \(\mu_{\bm{\beta}_m}(t)\) at
\(\mathbf{t}^\ast\) and
\(C_{\bm{\theta}_m}(\mathbf{t}^\ast, \mathbf{t}^\ast))\) is the
covariance matrix generated by evaluating the prior covariance function
\(C_{\bm{\theta}_m}(s,t)\) at
\(\mathbf{t}^\ast \times \mathbf{t}^\ast\). Using the properties of
multivariate normal distributions, the posterior distribution
\(d_m(\mathbf{t}^\ast) \mid \mathcal{D}_m, \ifthenelse{\value{bbsym4theta}=0}{\text{\faBasketballBall}}{\mathbf{\Theta}}_m\)
is also multivariate normal. This facilitates Bayesian estimation of the
distribution of the latent process governing the score difference given
the observed data from each match.

In addition to obtaining inference for the latent process we may also
estimate its time dynamics. This follows since a Gaussian process along
with its time derivatives (provided they exist) are distributed as a
multivariate Gaussian process (Cramer and Leadbetter (1967)). We may
therefore augment the hierarchical model in Equation
(\ref{eq:hierarchicalmodel}) with an additional latent structure of the
first and second derivatives of \(d_m\) with respect to time as
\begin{align}
  \begin{bmatrix}d_m(s)\\ d^{\prime}_m(t)\\ d^{\prime\prime}_m(u)\end{bmatrix} \mid \ifthenelse{\value{bbsym4theta}=0}{\text{\faBasketballBall}}{\mathbf{\Theta}}_m &\sim \mathcal{GP}\left(\begin{bmatrix}\mu_{\bm{\beta}_m}(s)\\ \mu^{\prime}_{\bm{\beta}_m}(t)\\ \mu^{\prime\prime}_{\bm{\beta}_m}(u)\end{bmatrix}, \begin{bmatrix}C_{\bm{\theta}_m}(s, s^\prime) & \partial_2 C_{\bm{\theta}_m}(s, t) & \partial_2^2 C_{\bm{\theta}_m}(s, u)\\ \partial_1 C_{\bm{\theta}_m}(t, s) & \partial_1 \partial_2 C_{\bm{\theta}_m}(t, t^\prime) & \partial_1 \partial_2^2 C_{\bm{\theta}_m}(t, u)\\ \partial_1^2 C_{\bm{\theta}_m}(u, s) & \partial_1^2\partial_2 C_{\bm{\theta}_m}(u, t) & \partial_1^2 \partial_2^2 C_{\bm{\theta}_m}(u, u^\prime)\end{bmatrix}\right)
\label{eq:dynamics}
\end{align} where \(^\prime\) and \(^{\prime\prime}\) denote the first
and second time derivatives and \(\partial_j^k\) is the \(k\)'th order
partial derivative with respect to the \(j\)'th variable. Combining the
models in Equations (\ref{eq:hierarchicalmodel}) and (\ref{eq:dynamics})
we obtain explicit expressions for the posterior distributions
\(d^\prime_m \mid \mathcal{D}_m, \ifthenelse{\value{bbsym4theta}=0}{\text{\faBasketballBall}}{\mathbf{\Theta}}_m\)
and
\(d^{\prime \prime}_m \mid \mathcal{D}_m, \ifthenelse{\value{bbsym4theta}=0}{\text{\faBasketballBall}}{\mathbf{\Theta}}_m\).
Specifically, the posterior joint distributions of the latent processes
is the following multivariate Gaussian process \begin{align}
\begin{bmatrix}d_m(s)\\ d_m^\prime(t)\\ d_m^{\prime\prime}(u)\end{bmatrix} \mid \mathcal{D}_m, \ifthenelse{\value{bbsym4theta}=0}{\text{\faBasketballBall}}{\mathbf{\Theta}}_m \sim \mathcal{GP}\left(\begin{bmatrix}\mu_{d_m}(s)\\ \mu_{d_m^\prime}(t)\\ \mu_{d_m^{\prime\prime}}(u)\end{bmatrix}, \begin{bmatrix} \Sigma_{d_m}(s, s^\prime) & \Sigma_{d_m d_m^\prime}(s,t) & \Sigma_{d_m d_m^{\prime\prime}}(s,u)\\ \Sigma_{d_m^\prime d_m}(t,s) & \Sigma_{d_m^\prime}(t,t^\prime) & \Sigma_{d_m^\prime d_m^{\prime\prime}}(t,u)\\ \Sigma_{d_m^{\prime\prime} d_m}(u,s) & \Sigma_{d_m^{\prime\prime} d_m^\prime}(u,t) & \Sigma_{d_m^{\prime\prime}}(u,u^\prime)\end{bmatrix}\right)
\label{eq:posterior}
\end{align} where explicit expressions for the posterior mean and
covariance functions are given in the online Supplementary Material.
Consequently, we can sample from this posterior joint distribution at
any finite number of time points as it corresponds to sampling from a
certain high-dimensional normal distribution. We utilize the posterior
samples of the first and second time derivatives of the latent process
to characterize the dynamical properties of each match through the Trend
Direction Index and the Excitement Trend Index.

We define the Trend Direction Index (TDI) of a particular match \(m\) as
the local posterior probability that \(d_m\) is an increasing function
at any time point \(t \in \mathcal{I}_m\). Under our model this is equal
to \begin{align}
\begin{split}
  \textrm{TDI}_m(t \mid \ifthenelse{\value{bbsym4theta}=0}{\text{\faBasketballBall}}{\mathbf{\Theta}}_m) &= P\left(d_m^\prime(t) > 0 \mid \mathcal{D}_m, \ifthenelse{\value{bbsym4theta}=0}{\text{\faBasketballBall}}{\mathbf{\Theta}}_m\right)\\
     &= \frac{1}{2} + \frac{1}{2}\mathop{\mathrm{Erf}}\left(\frac{\mu_{d_m^\prime}(t)}{2^{1/2}\Sigma_{d_m^\prime}(t, t)^{1/2}}\right)
\end{split}
\label{eq:TDIdef}
\end{align} where
\(\mathop{\mathrm{Erf}}\colon\, x \mapsto 2\pi^{-1/2}\int_0^x \exp(-u^2)\mathrm{d}u\)
is the error function and \(\mu_{d_m^\prime}\), and
\(\Sigma_{d_m^\prime}\) are the posterior mean and covariance functions
of the time derivative defined in Equation (\ref{eq:posterior}). The
interpretation of the TDI is that it quantifies the probability that one
team is currently increasing the differences in scores or equivalently
that they are changing the trend in their favor. A TDI equal to \(50\%\)
means that the game is in a stagnant state. We note that the TDI is
symmetric with respect to the reference team in the definition of the
score difference. If the reference team is switched, then the
\(\textrm{TDI}\) changes to \(1 - \textrm{TDI}\).

For each match we assign its Excitement Trend Index, \(\text{ETI}_m\),
as a global measure of game excitement. The index is defined as the
expected number of changes in monotonicity of the posterior distribution
of \(d_m\) which is equivalent to the expected number of zero-crossings
of the posterior distribution of \(d^\prime_m\). We hence define
\begin{align}
\begin{split}
  \text{ETI}_m \mid \ifthenelse{\value{bbsym4theta}=0}{\text{\faBasketballBall}}{\mathbf{\Theta}}_m &= \mathop{\mathrm{E}}\left[\#\left\{t \in \mathcal{I}_m : d^\prime_m(t) = 0\right\} \mid \mathcal{D}_m, \ifthenelse{\value{bbsym4theta}=0}{\text{\faBasketballBall}}{\mathbf{\Theta}}_m\right]\\
  &= \int_{\mathcal{I}_m} \int_{-\infty}^\infty \left|v\right| f_{d^\prime_m(t), d^{\prime\prime}_m(t)}(0, v \mid \mathcal{D}_m, \ifthenelse{\value{bbsym4theta}=0}{\text{\faBasketballBall}}{\mathbf{\Theta}}_m)\textrm{d}v\textrm{d}t \\
  &= \int_{\mathcal{I}_m} d\mathrm{ETI}_m(t \mid \ifthenelse{\value{bbsym4theta}=0}{\text{\faBasketballBall}}{\mathbf{\Theta}}_m)\mathrm{d}t
\end{split}
\label{eq:ETIdef}
\end{align} where
\(f_{d^\prime_m(t), d^{\prime\prime}_m(t)}(\cdot, \cdot \mid \mathcal{D}_m, \ifthenelse{\value{bbsym4theta}=0}{\text{\faBasketballBall}}{\mathbf{\Theta}}_m)\)
denotes the posterior density function of
\((d^\prime, d^{\prime\prime})\) at time \(t\) according to Equation
(\ref{eq:posterior}), and \(d\mathrm{ETI}_m\) is the instantaneous
posterior probability of a zero-crossing of \(d^\prime\) at any time
point \(t \in \mathcal{I}_m\). Integrating the instantaneous posterior
probability over the duration of a match gives us the ETI. Using
Equation (\ref{eq:posterior}) it can be shown that the instantaneous
posterior probability of a zero-crossing of \(d^\prime\) is equal to
\begin{align*}
d\mathrm{ETI}_m(t \mid \ifthenelse{\value{bbsym4theta}=0}{\text{\faBasketballBall}}{\mathbf{\Theta}}_m) = \lambda_m(t)\phi\left(\frac{\mu_{d_m^\prime}(t)}{\Sigma_{d_m^\prime}(t,t)^{1/2}}\right)\left(2\phi\left(\zeta_m(t)\right) + \zeta_m(t)\mathop{\mathrm{Erf}}\left(\frac{\zeta_m(t)}{2^{1/2}}\right)\right)
\end{align*} where
\(\phi\colon\, x \mapsto 2^{-1/2}\pi^{-1/2}\exp(-\frac{1}{2}x^2)\) is
the standard normal density function, and \(\lambda_m\), \(\omega_m\)
and \(\zeta_m\) are defined as \begin{gather*}
  \lambda_m(t) = \frac{\Sigma_{d_m^{\prime\prime}}(t,t)^{1/2}}{\Sigma_{d_m^\prime}(t,t)^{1/2}}\left(1-\omega_m(t)^2\right)^{1/2}, \quad \omega_m(t) = \frac{\Sigma_{d_m^\prime d_m^{\prime\prime}}(t,t)}{\Sigma_{d_m^\prime}(t,t)^{1/2}\Sigma_{d_m^{\prime\prime}}(t,t)^{1/2}}\\
  \zeta_m(t) = \frac{\mu_{d_m^\prime}(t)\Sigma_{d_m^{\prime^\prime}}(t,t)^{1/2}\omega_m(t)\Sigma_{d_m^\prime}(t,t)^{-1/2} - \mu_{d_m^{\prime\prime}}(t)}{\Sigma_{d_m^{\prime\prime}}(t,t)^{1/2}\left(1 - \omega_m(t)^2\right)^{1/2}}
\end{gather*} The derivation of the expression of \(d\mathrm{ETI}_m\)
can be found in the online Supplementary Material to Jensen and Ekstrøm
(2020b). While no closed-form expression for
\(\text{ETI}_m \mid \ifthenelse{\value{bbsym4theta}=0}{\text{\faBasketballBall}}{\mathbf{\Theta}}_m\)
seems to exist, the integration can be performed numerically. We note
that the ETI is also invariant with respect to the choice of reference
team in the definition of the score differences as it is defined as the
expected number of both up- and down-crossings at zero of the posterior
trend.

Both TDI and ETI as defined in Equations (\ref{eq:TDIdef}) and
(\ref{eq:ETIdef}) are random variables due to their dependence on the
hyper-parameters
\(\ifthenelse{\value{bbsym4theta}=0}{\text{\faBasketballBall}}{\mathbf{\Theta}}_m\).
In our Bayesian framework these are specified under an additional layer
of prior distributions according to
\(H(\ifthenelse{\value{bbsym4theta}=0}{\text{\faBasketballBall}}{\mathbf{\Theta}}_m \mid \bm{\Psi}_m)\).
By fitting the model using Markov-Chain Monte Carlo methods (MCMC) we
obtain samples from the posterior distribution
\(\widetilde{\ifthenelse{\value{bbsym4theta}=0}{\text{\faBasketballBall}}{\mathbf{\Theta}}}_m \sim P(\ifthenelse{\value{bbsym4theta}=0}{\text{\faBasketballBall}}{\mathbf{\Theta}}_m \mid \mathcal{D}_m, \bm{\Psi}_m, \mathbf{t}_m)\)
and the posterior estimates of TDI and ETI are therefore the random
variables
\(\textrm{TDI}_m(t \mid \widetilde{\ifthenelse{\value{bbsym4theta}=0}{\text{\faBasketballBall}}{\mathbf{\Theta}}}_m)\)
and
\(\textrm{ETI}_m \mid \widetilde{\ifthenelse{\value{bbsym4theta}=0}{\text{\faBasketballBall}}{\mathbf{\Theta}}}_m\).

\hypertarget{sec:estimation}{%
\section{Estimation}\label{sec:estimation}}

A completion of the model in Equation (\ref{eq:hierarchicalmodel})
requires a specification of the prior mean and covariance functions for
the latent process. The choice of these are application specific and can
be based on prior knowledge of the game dynamics. We refer to the
discussion in Jensen and Ekstrøm (2020b) for more information on such
choices.

In our application we used a constant prior mean and the squared
exponential covariance function given by \begin{align*}
  \mu_{\bm{\beta}_m}(t) = \beta_m, \quad C_{\bm{\theta}_m}(s, t) = \alpha^2_m\exp\left(-\frac{(s-t)^2}{2\rho^2_m}\right)
\end{align*} and thus
\(\ifthenelse{\value{bbsym4theta}=0}{\text{\faBasketballBall}}{\mathbf{\Theta}}_m = (\beta_m,\alpha_m, \rho_m, \sigma_m) \in \mathbb{R} \times \mathbb{R}_{>0}^3\).
These assumptions ensure well-defined and infinitely differentiable
sample paths of \(d_m\). For the hyper-parameters
\(\ifthenelse{\value{bbsym4theta}=0}{\text{\faBasketballBall}}{\mathbf{\Theta}}_m\)
we used independent, heavy-tailed distribution with a moderate variance
centered at the marginal maximum likelihood estimates of the form
\begin{align*}
H(\ifthenelse{\value{bbsym4theta}=0}{\text{\faBasketballBall}}{\mathbf{\Theta}}_m \mid \bm{\Psi}_m) = H(\beta_m \mid \Psi_{\beta_m})H(\alpha_m \mid \Psi_{\alpha_m})H(\rho_m \mid \Psi_{\rho_m})H(\sigma_m \mid \Psi_{\sigma_m})
\end{align*} with \begin{align*}
\beta_{m} \sim T_4\left(\widehat{\beta_m^\text{ML}}, 5\right), \; \alpha_m \sim T^+_4\left(\widehat{\alpha_m^\text{ML}}, 5\right), \; \rho_m \sim T_4^+\left(\widehat{\rho_m^\text{ML}}, 5\right), \; \sigma_m \sim T^+_4\left(\widehat{\sigma_m^\text{ML}}, 5\right)
\end{align*} where \(T_{\text{df}}\) denotes a location-scale T
distribution with \(\mathrm{df}\) degrees of freedom,
\(T_{\text{df}}^+\) denotes the same distribution but truncated to the
positive real line, and \(\widehat{\,_m^\text{ML}}\) notes the marginal
maximum likelihood estimate of the corresponding hyper-parameter. By the
properties of the model in Equation (\ref{eq:hierarchicalmodel}) the
marginal maximum log-likelihood function has the following closed-form
expression \begin{align*}
\log L(\ifthenelse{\value{bbsym4theta}=0}{\text{\faBasketballBall}}{\mathbf{\Theta}}_m \mid \mathcal{D}_m) &\propto - \frac{1}{2}\log |C_{\bm{\theta}_m}(\mathbf{t}_m, \mathbf{t}_m) + \sigma_m^2 I|\\
&- \frac{1}{2}(\mathbf{D}_m - \mu_{\bm{\beta}_m}(\mathbf{t}_m))^T\left[C_{\bm{\theta}_m}(\mathbf{t}_m, \mathbf{t}_m) + \sigma_m^2 I\right]^{-1}(\mathbf{D}_m - \mu_{\bm{\beta}_m}(\mathbf{t}_m))
\end{align*} and the marginal maximum likelihood estimates
\(\widehat{\ifthenelse{\value{bbsym4theta}=0}{\text{\faBasketballBall}}{\mathbf{\Theta}}_m^\text{ML}} = \mathop{\mathrm{arg\,sup}}_{\ifthenelse{\value{bbsym4theta}=0}{\text{\faBasketballBall}}{\mathbf{\Theta}}} \log L(\ifthenelse{\value{bbsym4theta}=0}{\text{\faBasketballBall}}{\mathbf{\Theta}}\mid \mathcal{D}_m)\)
can obtained by numerical optimization or as the roots to the score
equations
\(\nabla_{\ifthenelse{\value{bbsym4theta}=0}{\text{\faBasketballBall}}{\mathbf{\Theta}}} \log L(\ifthenelse{\value{bbsym4theta}=0}{\text{\faBasketballBall}}{\mathbf{\Theta}}\mid \mathcal{D}_m) = 0\).

We have implemented our model in the probabilistic programming language
Stan (Carpenter et al. 2017) that uses a Hamiltonian Markov Chain Monte
Carlo algorithm to sample from the distributions of
\(\textrm{TDI}_m(t \mid \widetilde{\ifthenelse{\value{bbsym4theta}=0}{\text{\faBasketballBall}}{\mathbf{\Theta}}}_m)\)
and
\(d\mathrm{ETI}_m(t \mid \widetilde{\ifthenelse{\value{bbsym4theta}=0}{\text{\faBasketballBall}}{\mathbf{\Theta}}}_m)\)
on any finite set of time points in \(\mathcal{I}_m\). Posterior summary
measures of
\(\textrm{TDI}_m(t \mid \widetilde{\ifthenelse{\value{bbsym4theta}=0}{\text{\faBasketballBall}}{\mathbf{\Theta}}}_m)\)
and
\(\mathrm{ETI}_m \mid \widetilde{\ifthenelse{\value{bbsym4theta}=0}{\text{\faBasketballBall}}{\mathbf{\Theta}}}_m\)
can then be calculated and reported from the posterior samples as e.g.,
the mean or median along with \((1-\alpha)100\%\) credible intervals.

\hypertarget{sec:application}{%
\section{Application: The 2019--2020 NBA basketball
season}\label{sec:application}}

To illustrate the applicability of our proposed methodology we apply it
to data from all regular games from the 2019--2020 NBA basketball
season. The data was obtained from Sports Reference LLC (2020) and is
provided in Jensen and Ekstrøm (2020a). A lot of points are scored
during a basketball match so it is easy to see the development of the
score difference in a single match.

The 2019--2020 NBA season was suspended mid-March due to COVID-19 but it
was resumed again in July 2020. There were a total of 1059 regular
season matches. The subsequent playoffs comprised 84 matches including
the final for a grand total of 1143 matches. For ease of comparison we
are only considering the first 48 regular minutes of each match --- any
part of a match that goes into overtime will be disregarded, and we
hence let \(\mathcal{I}_m = [0; 48]\) minutes.

We wish to use the trend analysis proposed for three purposes: 1) to
show how the TDI can be used to infer real-time and post-game evaluation
of the trends in a match. 2) To evaluate the ETI for all 1143 matches in
the 2019--2020 season to provide background and reference information
about the matches, and 3) to summarize ETI at the team level in order to
identify groups of teams more/less likely to give an exciting game.

For each match we used the prior specification from Section
\ref{sec:estimation}. We ran four independent chains for 50,000
iterations each with half of the iterations used for warm-up and
evaluated the posterior distribution of
\(\textrm{TDI}_m(t \mid \widetilde{\ifthenelse{\value{bbsym4theta}=0}{\text{\faBasketballBall}}{\mathbf{\Theta}}}_m)\)
and
\(d\mathrm{ETI}_m(t \mid \widetilde{\ifthenelse{\value{bbsym4theta}=0}{\text{\faBasketballBall}}{\mathbf{\Theta}}}_m)\)
on an equidistant grid of 241 time points in \([0; 48]\). The posterior
distribution of
\(\text{ETI}_m \mid \widetilde{\ifthenelse{\value{bbsym4theta}=0}{\text{\faBasketballBall}}{\mathbf{\Theta}}}_m\)
was calculated by numerical integration using the trapezoidal method.

\begin{figure}[htbp]
\includegraphics{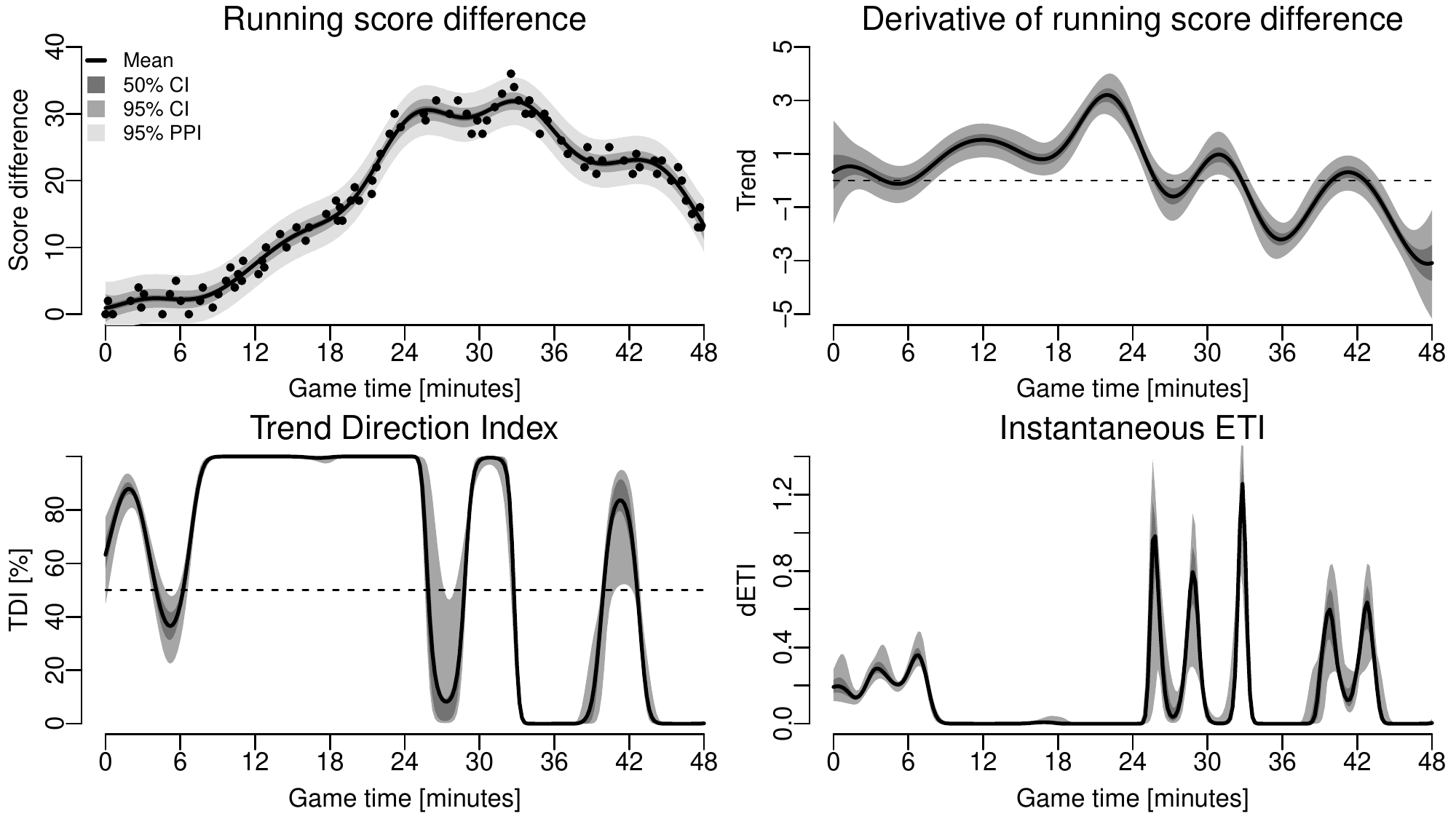}
\caption{Results from fitting the latent Gaussian Process to the final match between LA Lakers and Miami Heat in the 2019--2020 NBA season. Larger values in running score differences and the Trend Direction Index reflect the situation where LA Lakers are doing better. The top left panel shows the posterior distribution of the the latent process with the posterior means in bold. The gray areas show point-wise credible and posterior prediction intervals. The top right panel provides similar information for the the posterior trend. The bottom left panel shows the TDI and can be used to read off probability statements about the trends in the running score difference. The bottom right panel shows the local ETI and quantifies the instantaneous probability of a change in monotonicity of the score differences.}
\label{fig1b}
\end{figure}

For our first purpose of analyzing trends in individuals matches we
consider the final match of the 2019-2020 season. The raw data for the
running score difference between LA Lakers and Miami Heat was shown in
Figure \ref{fig1}. Figure \ref{fig1b} shows the results from the
post-game analysis.

Evaluating the game trends from the TDI in Figure \ref{fig1b} shows that
LA Lakers had control of most of the match since the posterior
probability of a positive trend was high throughout most of the match.
Only towards the end of the third quarter did Miami Heat gain the upper
hand and had a period where they fought back. In the 4th quarter from
around 38 minutes to 43 minutes we can see that mean TDI increases to
over 80\% but the probability interval is very wide reflecting that it
is difficult to say whether the trend is increasing or if it might as
well just be random fluctuations in scores. Similarly for the first half
of the 3rd quarter. Teams wishing to evaluate the match should primarily
concentrate on periods where the latent trend and its probability
interval is either close to 50\% or when the trend is disadvantageous
for the team. The spikes observed in the local ETI (lower right plot of
Figure \ref{fig1b}) indicate the time points where the monotonicity of
the underlying trend is changing sign, with higher values of \(d\)ETI
representing more steep changes.

\begin{figure}[htbp]
\includegraphics{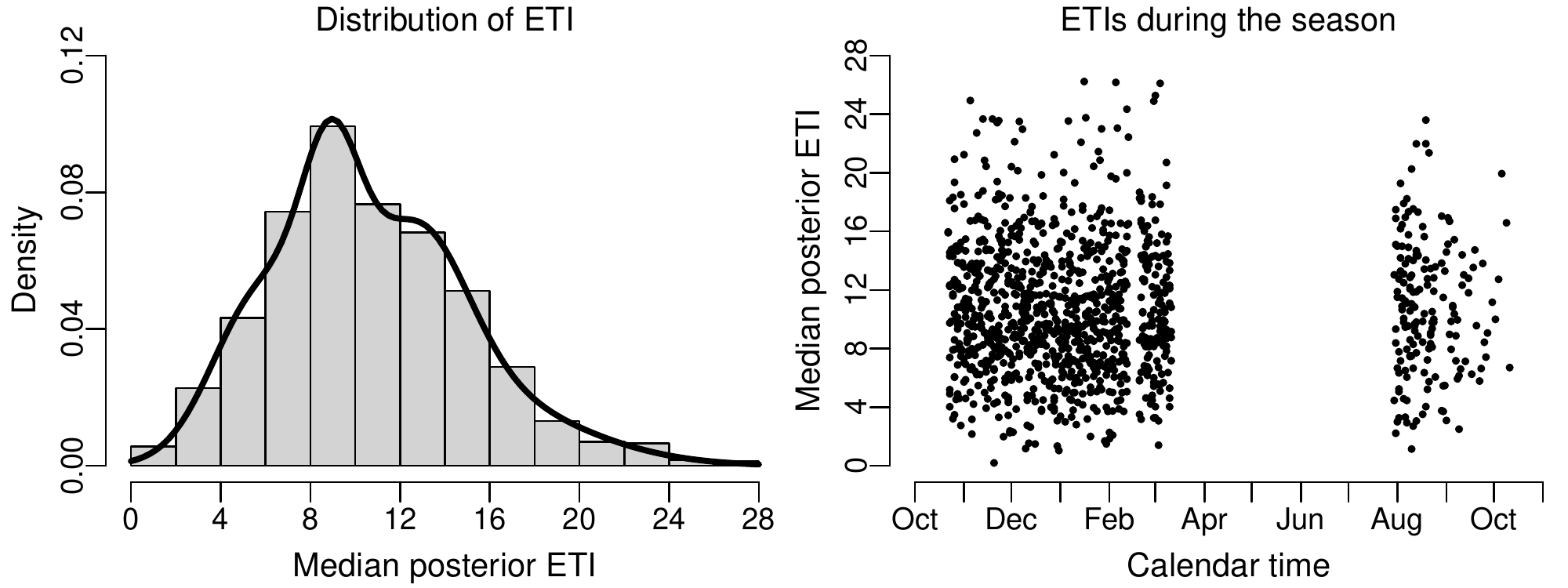}
\caption{Histogram and Gaussian mixture model estimate of the distribution of the 1143 median posterior ETIs from the NBA 2019--2020 season (left panel) and the median posterior ETIs as a function of calendar time from October 22, 2019 to October 11, 2020 (right panel).}
\label{fig2}
\end{figure}

To evaluate the overall distribution of the ETIs we fitted our model to
each of the 1143 matches during the season and estimated the ETI for
each. The results are summarized in the left panel of Figure \ref{fig2}
showing the marginal distribution of the median posterior ETIs. The
solid line shows the fit of a Gaussian mixture model with four
components, where the number of components were determined by sequential
bootstrapped likelihood ratio tests (Scrucca et al. (2016)). The
marginal distribution of the posterior median ETIs is right skewed
(skewness = 0.52) with a range of {[}0.19; 26.23{]}, and an median of
10.12 (mean = 10.64, SD = 4.51). This implies that the time-varying
score differences of the games in the season changes monotonicity
approximately 10 times during a game on average but with a large
variation between matches. For comparison, the final match between LA
Lakers and Miami Heat shown in Figure \ref{fig1b} has an median
posterior ETI of \(6.71\).

We wished to examine if there was a calendar time effect on game
excitement as the season progressed in order to investigate if we would
find that games became more exciting as the teams fought to stay in the
competition to enter the final playoffs or if we could detect fatigue
over the season. The right panel of Figure \ref{fig2} shows the median
posterior ETIs as a function of calendar time at which the matches were
played. Besides illustrating the gap from the COVID-19 hiatus, the
figure shows that the excitement indices are relatively evenly
distributed throughout the season.

\begin{figure}[htbp]
\includegraphics{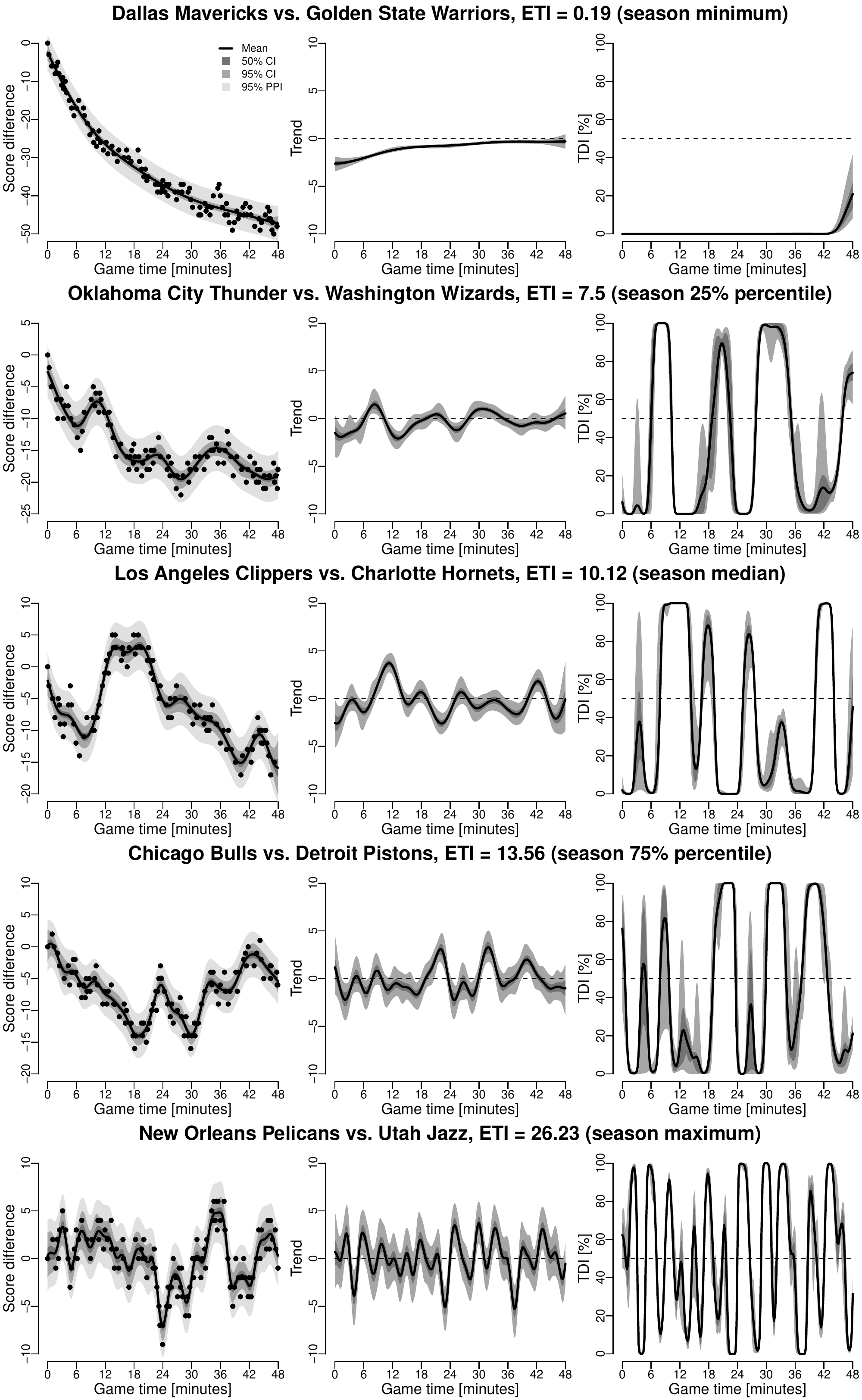}
\caption{Observed score differences with posteriors of the latent processes (left panels), posterior trends (middle panels) and posterior Trend Direction Indices (right panels) for five games from the NBA season 2019--2020 with median posterior Excitement Trend Indices corresponding to the 0\%, 25\%, 50\%, 75\%, and 100\% percentiles of the distribution of all games in the season. Gray regions depict 50\% and 95\% point-wise credible intervals and 95\% posterior prediction intervals.}
\label{fig3}
\end{figure}

When matches are ranked from lowest to highest median posterior ETI, we
can extract the individual analyses for matches representing the full
span of the ETI range. Figure \ref{fig3} shows the analysis results of
our proposed method for the matches with minimum, 1st, 2nd, 3rd
quantile, and maximum median posterior ETI. It is clear from the
observed running score differences, posterior trends, and the TDIs that
these five matches represent substantially different game experiences.

The first row of Figure \ref{fig3} for Dallas Mavericks vs the Golden
State Warriors show a very one-sided match leading to the minimum ETI
during the season. The third row of Figure \ref{fig3} for the LA
Clippers vs Charlotte Hornets match indicates that while the Hornets did
lead throughout most of first quarter, the Clippers reversed the game
towards the end of first quarter and kept the lead through most of 2nd
quarter. After that the Hornets started to keep the lead and the
Clippers never managed to make a proper comeback and the game trend was
rather flat in the last two quarters of the game since the two teams
more or less kept the pace with each other except for the effort shown
midway in quarter four. In contrast, the match between New Orleans
Pelicans vs Utah Jazz (5th row in Figure \ref{fig3}) showed trends that
varied direction frequently and where the TDI showed alternating periods
of scoring bursts making it a very exciting and unpredictable game. The
online Supplementary Material lists summary statistics of the estimated
ETIs for all 1143 matches.

Summarizing the median posterior ETIs at the team level across all
matches during the season lead to comparable values for all 30 teams.
Table \ref{tab:teamsummary} shows the summary statistics for all 30
teams ordered by their season average excitement. The New Orleans
Pelicans had the highest median posterior ETI averaged over the season
with an average median posterior ETI of 11.67 (SD = 4.91, IQR =
\([3.10; 23.54]\)), while the Charlotte Hornets had the lowest average
median posterior ETI with a value of 9.26 (SD = 3.74, IQR =
\([1.96; 15.95]\)). The small fluctuation of the averages suggests that
the teams are comparable in terms of excitement when averaging across
all their games during the season, and that the major source of
variation in excitement during the season (as seen in Figure \ref{fig2})
is governed by the specific matches.

Although the team averages in Table \ref{tab:teamsummary} show limited
variability it is of interest to estimate a number of subgroups among
the teams that exhibited similar degree of excitement on average during
the season -- effectively clustering the teams. This would enable fans,
promoters, and sponsors to infer which teams were more likely to partake
in an exciting game. The problem is mathematically equivalent to looking
at the relationship between the median posterior ETIs as the outcome in
a linear regression model where the explanatory categorical variable
ranges over the set of all partitions of the 30 teams, and as the
objective we seek the smallest number of partitions that best explains
the observed outcome by comparing all possible splits of the ranked
teams for a given number of partitions. This will then define subgroups
of teams.

As the optimization criterion for the subgroup identification we used
the root mean squared error of prediction based on leave-one-out
cross-validation, denoted \(\textrm{RMSEP}_\text{LOO-CV}^{C = c}\) where
\(c\) is the number of subgroups. Our sequential optimization procedure
showed that the optimization criterion stabilized at four subgroups:
\(\textrm{RMSEP}_\text{LOO-CV}^{C=2} = 4.493\),
\(\textrm{RMSEP}_\text{LOO-CV}^{C=3} = 4.49\),
\(\textrm{RMSEP}_\text{LOO-CV}^{C=4} = 4.489\), and subsequently for
\(C = 5,\ldots,8\) it remained at the same value. The labels of these
groups are shown in the rightmost column in Table \ref{tab:teamsummary}.
The result is thus an identification of three change-points in the
ranking of the teams according to median posterior ETI averaged across
the season. The most noticeable result is that Charlotte Hornets
constitute a singleton since that team has substantial lower ETI than
the team with the second lowest ETI. To maximize the probability of
seeing an exciting game it would thus have be wise to avoid matches in
which the Hornets were playing.

\begin{table}

\caption{\label{tab:unnamed-chunk-1}Team specific median posterior Excitement Trend Indices summarized across all their matches in the 2019--2020 NBA season and ordered by the season average. The group column shows the clustering induced by our optimization procedure.\label{tab:teamsummary}}
\centering
\begin{tabular}[t]{lrr|rrr|r}
\toprule
  & Average & SD & 2.5\% & 50\% & 97.5\% & Group\\
\midrule
New Orleans Pelicans & 11.67 & 4.91 & 3.10 & 11.59 & 23.54 & A\\
Washington Wizards & 11.36 & 4.46 & 2.98 & 11.63 & 18.28 & A\\
San Antonio Spurs & 11.31 & 5.31 & 3.52 & 9.88 & 23.44 & A\\
Oklahoma City Thunder & 11.26 & 5.32 & 1.68 & 11.55 & 23.74 & A\\
Portland Trail Blazers & 11.21 & 4.46 & 3.31 & 10.82 & 18.48 & A\\
Los Angeles Lakers & 11.16 & 4.43 & 3.61 & 10.52 & 19.79 & A\\
Philadelphia 76ers & 11.09 & 4.70 & 2.73 & 10.28 & 21.53 & A\\
Minnesota Timberwolves & 11.09 & 4.28 & 4.01 & 10.69 & 19.46 & A\\
Utah Jazz & 10.96 & 5.44 & 3.09 & 9.76 & 23.54 & A\\
Memphis Grizzlies & 10.94 & 4.59 & 3.04 & 10.36 & 20.84 & A\\
Cleveland Cavaliers & 10.90 & 4.50 & 4.93 & 10.39 & 22.36 & A\\
\hline
Toronto Raptors & 10.69 & 4.44 & 3.48 & 10.33 & 21.79 & B\\
Houston Rockets & 10.67 & 4.12 & 3.52 & 10.74 & 19.17 & B\\
Chicago Bulls & 10.66 & 3.98 & 2.95 & 10.65 & 17.49 & B\\
Boston Celtics & 10.61 & 4.66 & 2.81 & 10.19 & 20.92 & B\\
Orlando Magic & 10.56 & 4.89 & 2.72 & 10.41 & 21.01 & B\\
Milwaukee Bucks & 10.55 & 4.36 & 2.77 & 9.59 & 18.13 & B\\
Los Angeles Clippers & 10.54 & 4.14 & 4.05 & 10.72 & 20.56 & B\\
Miami Heat & 10.44 & 4.29 & 3.28 & 9.91 & 18.54 & B\\
Golden State Warriors & 10.40 & 4.58 & 3.37 & 10.24 & 18.72 & B\\
Indiana Pacers & 10.37 & 3.99 & 3.31 & 10.22 & 18.54 & B\\
\hline
Brooklyn Nets & 10.33 & 4.57 & 3.07 & 10.06 & 18.89 & C\\
Dallas Mavericks & 10.27 & 4.99 & 3.00 & 9.04 & 22.09 & C\\
Atlanta Hawks & 10.26 & 4.35 & 2.50 & 9.75 & 18.31 & C\\
Phoenix Suns & 10.25 & 4.32 & 3.78 & 9.29 & 19.81 & C\\
Denver Nuggets & 10.16 & 4.39 & 3.42 & 9.59 & 19.61 & C\\
New York Knicks & 10.11 & 4.05 & 2.84 & 9.87 & 18.47 & C\\
Sacramento Kings & 9.94 & 4.19 & 3.66 & 9.30 & 18.93 & C\\
Detroit Pistons & 9.86 & 4.00 & 3.73 & 9.44 & 17.08 & C\\
\hline
Charlotte Hornets & 9.26 & 3.74 & 1.96 & 9.22 & 15.95 & D\\
\bottomrule
\end{tabular}
\end{table}

\begin{figure}[htbp]
\includegraphics{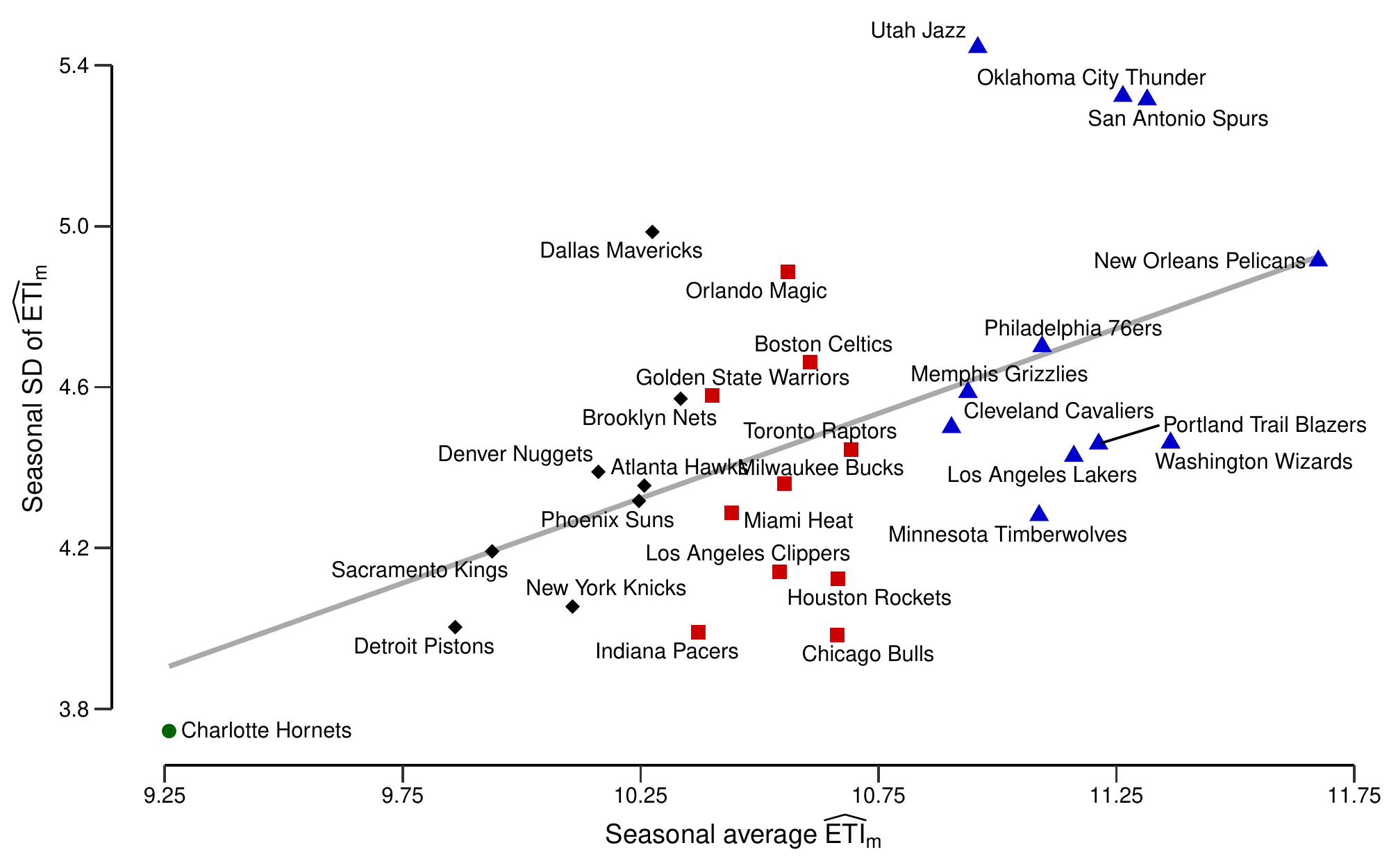}
\caption{Seasonal SD of the estimated ETIs plotted against the average estimated ETI for each team. Shapes/colors indicate ETI groups. The line represents the best linear regression line through the origin. The plot suggests a linear relationship between the SDs and averages indicating a fairly constant coefficient of variation. Teams with higher excitement trend index are more likely to have higher variation in their excitement scores.}
\label{fig:linplot}
\end{figure}

Figure \ref{fig:linplot} shows the estimated linear association between
the seasonal standard deviations and average of the median posterior
ETIs for each team. The teams follow a straight line through the origin
fairly well, suggesting that the coefficient of variation is fairly
constant across the teams. This also suggests that teams with large
seasonal average excitement are also more likely to be part of really
spectacular matches. Teams consistently contributing with highly
exciting matches during the season should therefore be represented in
the bottom right part of the plot. Assessing which teams that where
generally most exiting to watch during the season is thus a
bias-variance trade-off decision. While the New Orleans Pelicans were
most exciting on average, the Washington Wizards where second-most
exciting on average but also had a notably lower seasonal standard
deviation. This means that although the latter team was less exciting on
average during the season, their level of excitement was more stable
throughout. On the other hand, the San Antonio Spurs had the third
largest average excitement but simultaneously also a much larger
standard deviation making the excitement of their individual matches
much less reliable throughout the season.

\hypertarget{sec:discussion}{%
\section{Discussion}\label{sec:discussion}}

We have introduced the Trend Direction Index as a measure to estimate
and evaluate the trends in running score difference in sports. The Trend
Direction Index is based on a latent Gaussian process model and enables
us to make Bayesian in-game and post-game evaluations about the
underlying trends in scoring patterns and to attach easily interpretable
probability statements to the results. In addition we have presented the
Excitement Trend Index as the expected number of monotonicity changes in
the underlying trend and we showed how it can be used to gauge how
exciting a match will be: if one team is consistently outperforming the
other then the match quickly becomes one-sided and less exciting. Both
indices have intuitive interpretations that are easily conveyed to
non-statisticians, coaches, players, and commentators.

We have showed how the proposed method can be used to analyze single
matches in order to determine strategies to identify periods throughout
the game where the momentum of the game changes. The model utilizing the
latent trend enables a highly detailed modeling approach where game
development can be followed from minute to minute. This will facilitate
and improve post-game coaching and influence future game tactics.

Our analysis of all matches in the 2019--2020 NBA season showed that
different values of the ETI captured games with vastly different
features and showed that it could be used as a tool to discriminate the
teams. In this case, the latent Gaussian process benefits greatly from
the large numbers of scorings that are typical in basketball, since that
means that there will be a large and frequent number of observations
within each match that the model can utilize. In contrast, sports with
few scorings such as soccer may prove a more difficult task simply
because there is very few changes in the running score throughout a
match.

There are a couple of future research ideas that could extend our
current approach of using the latent Gaussian trend to infer measures
for game excitement. One idea is to define a weighted version of the
Excitement Trend Index, \(\mathrm{wETI}_m\), so that changes in
monotonicity of the score differences are i) weighted higher towards the
end of the game and ii) weighted lower if one team is already far away
of the other team as measured by the absolute value of the posterior
mean \(\mu_{d_m}\). This motivates a modification of the definition of
the ETI in Equation (\ref{eq:ETIdef}) to the following weighted form
\begin{align*}
  \mathrm{wETI}_m\mid \ifthenelse{\value{bbsym4theta}=0}{\text{\faBasketballBall}}{\mathbf{\Theta}}_m = \int_{\mathcal{I}_m} d\mathrm{ETI}_m(t \mid \ifthenelse{\value{bbsym4theta}=0}{\text{\faBasketballBall}}{\mathbf{\Theta}}_m)w(t, |\mu_{d_m}(t)|)\mathrm{d}t
\end{align*} where
\(w\colon\, \mathcal{I}_m \times \mathbb{R}_{\geq 0} \mapsto \mathbb{R}_{\geq 0}\)
is a weight function that is increasing in its first variable and
decreasing in its second variable. Such weight functions could be
constructed as a product of two kernel functions defined on their
individual domains and with bandwidths based on studies of psychological
perception.

Another approach for quantifying excitement would be to define it at the
team-level instead of at the match level. In that case one could define
team-specific Trend Excitement Indices nested with a match by looking at
both the up- and down-crossing of \(df_m\) at zero. This would result in
two excitement indices for each match,
\((\mathrm{ETI}_{am}, \mathrm{ETI}_{bm})\) for teams \(a\) and \(b\)
which would reflect how exciting each team were in match \(m\) with
respect to chancing the sign of the score differences in their favor.

In conclusion, we have provided an analytical framework for analyzing
the trend in the running score difference in sports matches. The latent
Gaussian process model requires very few assumptions which makes the
modeling approach very flexible and applicable to a multitude of sports.

\hypertarget{bibliography}{%
\section*{Bibliography}\label{bibliography}}
\addcontentsline{toc}{section}{Bibliography}

\hypertarget{refs}{}
\leavevmode\hypertarget{ref-baboota}{}%
Baboota, Rahul, and Harleen Kaur. 2018. ``Predictive Analysis and
Modelling Football Results Using Machine Learning Approach for English
Premier League.'' \emph{International Journal of Forecasting} 35
(March). \url{https://doi.org/10.1016/j.ijforecast.2018.01.003}.

\leavevmode\hypertarget{ref-carpenter2017stan}{}%
Carpenter, Bob, Andrew Gelman, Matthew D Hoffman, Daniel Lee, Ben
Goodrich, Michael Betancourt, Marcus Brubaker, Jiqiang Guo, Peter Li,
and Allen Riddell. 2017. ``Stan: A Probabilistic Programming Language.''
\emph{Journal of Statistical Software} 76 (1).

\leavevmode\hypertarget{ref-cattelan2013bradley}{}%
Cattelan, Manuela, Cristiano Varin, and David Firth. 2013. ``Dynamic
Bradley--Terry Modelling of Sports Tournaments.'' \emph{Journal of the
Royal Statistical Society: Series C (Applied Statistics)} 62 (1):
135--50.
\url{https://doi.org/https://doi.org/10.1111/j.1467-9876.2012.01046.x}.

\leavevmode\hypertarget{ref-chen2018functional}{}%
Chen, Tao, and Qingliang Fan. 2018. ``A Functional Data Approach to
Model Score Difference Process in Professional Basketball Games.''
\emph{Journal of Applied Statistics} 45 (1): 112--27.

\leavevmode\hypertarget{ref-chen2020rank}{}%
Chen, Yaqing, Matthew Dawson, and Hans-Georg Müller. 2020. ``Rank
Dynamics for Functional Data.'' \emph{Computational Statistics \& Data
Analysis}, 106963.

\leavevmode\hypertarget{ref-cramer1967stationary}{}%
Cramer, Harald, and M. R. Leadbetter. 1967. \emph{Stationary and Related
Stochastic Processes -- Sample Function Properties and Their
Applications.} John Wiley \& Sons, Inc.

\leavevmode\hypertarget{ref-ekstrom2020trps}{}%
Ekstrøm, Claus Thorn, Hans Van Eetvelde, Christophe Ley, and Ulf
Brefeld. 2020. ``Evaluating One-Shot Tournament Predictions.''
\emph{Journal of Sports Analytics} Preprint (Preprint): 1--10.
\url{https://doi.org/10.3233/JSA-200454}.

\leavevmode\hypertarget{ref-gabel2012random}{}%
Gabel, Alan, and Sidney Redner. 2012. ``Random Walk Picture of
Basketball Scoring.'' \emph{Journal of Quantitative Analysis in Sports}
8 (1).

\leavevmode\hypertarget{ref-groll2019}{}%
Groll, Andreas, Christophe Ley, Gunther Schauberger, and Hans Van
Eetvelde. 2019. ``A hybrid random forest to predict soccer matches in
international tournaments.'' \emph{Journal of Quantitative Analysis in
Sports} 15: 271--88.

\leavevmode\hypertarget{ref-Gu2019}{}%
Gu, Wei, and Thomas L. Saaty. 2019. ``Predicting the Outcome of a Tennis
Tournament: Based on Both Data and Judgments.'' \emph{Journal of Systems
Science and Systems Engineering} 28 (3): 317--43.
\url{https://doi.org/10.1007/s11518-018-5395-3}.

\leavevmode\hypertarget{ref-HavingABallGitHub}{}%
Jensen, Andreas Kryger, and Claus Thorn Ekstrøm. 2020a. ``GitHub
Repository for Having a Ball.'' 2020.
\url{https://github.com/aejensen/Having-a-Ball}.

\leavevmode\hypertarget{ref-ToT}{}%
---------. 2020b. ``Quantifying the Trendiness of Trends.''
\emph{Journal of the Royal Statistical Society: Series C}.

\leavevmode\hypertarget{ref-skellam2}{}%
Karlis, Dimitris, and Ioannis Ntzoufras. 2003. ``Analysis of Sports Data
by Using Bivariate Poisson Models.'' \emph{Journal of the Royal
Statistical Society: Series D (the Statistician)} 52 (3): 381--93.
\url{https://doi.org/10.1111/1467-9884.00366}.

\leavevmode\hypertarget{ref-rasmussen2003gaussian}{}%
Rasmussen, C. E., and C. K. I. Williams. 2006. \emph{Gaussian Processes
in Machine Learning}. MIT Press.

\leavevmode\hypertarget{ref-mclust}{}%
Scrucca, Luca, Michael Fop, T. Brendan Murphy, and Adrian E. Raftery.
2016. ``mclust 5: Clustering, Classification and Density Estimation
Using Gaussian Finite Mixture Models.'' \emph{The R Journal} 8 (1):
289--317. \url{https://doi.org/10.32614/RJ-2016-021}.

\leavevmode\hypertarget{ref-BBreference}{}%
Sports Reference LLC. 2020. ``Basketball Reference.'' 2020.
\url{https://www.basketball-reference.com/}.

\end{document}


\maketitle

\newcounter{bbsym4theta}
\setcounter{bbsym4theta}{0}

\hypertarget{expressions-for-the-posterior-distributions-of-d_m-dprime_m-dprime-prime_m}{%
\section{\texorpdfstring{Expressions for the posterior distributions of
\((d_m, d^\prime_m, d^{\prime \prime}_m)\)}{Expressions for the posterior distributions of (d\_m, d\^{}\textbackslash prime\_m, d\^{}\{\textbackslash prime \textbackslash prime\}\_m)}}\label{expressions-for-the-posterior-distributions-of-d_m-dprime_m-dprime-prime_m}}

The joint distribution of \((d, d^\prime, d_m^{\prime \prime})\)
conditional on the observed score differences
\(\mathcal{D}_m = (\mathbf{D}_m, \mathbf{t}_m)\) and the
hyper-parameters
\(\ifthenelse{\value{bbsym4theta}=0}{\text{\faBasketballBall}}{\mathbf{\Theta}}_m\)
evaluated at any finite vector \(\mathbf{t}^\ast\) of \(p\) time points
follows the multivariate normal distribution \begin{align*}
\begin{bmatrix}d_m(\mathbf{t}^\ast)\\ d_m^{\prime}(\mathbf{t}^\ast)\\ d_m^{\prime \prime}(\mathbf{t}^\ast)\end{bmatrix} \mid \mathcal{D}_m, \ifthenelse{\value{bbsym4theta}=0}{\text{\faBasketballBall}}{\mathbf{\Theta}}_m \sim N\left(\bm{\mu}_m,  \bm{\Sigma}_m\right)
\end{align*} where \(\bm{\mu}_m \in \mathbb{R}^{3p}\) is the column
vector of posterior means and
\(\bm{\Sigma}_m \in \mathbb{R}^{3p \times 3p}\) is the posterior
covariance matrix. These can be partitioned as \begin{align*}
  \bm{\mu}_m = \begin{bmatrix}\mu_{d_m}(\mathbf{t^\ast})\\ \mu_{d_m^\prime}(\mathbf{t^\ast})\\ \mu_{d_m^{\prime\prime}}(\mathbf{t^\ast})\end{bmatrix}, \quad \bm{\Sigma}_m = \begin{bmatrix}\Sigma_{d_m}(\mathbf{t^\ast},\mathbf{t^\ast}) &  \Sigma_{d_m d_m^\prime}(\mathbf{t^\ast},\mathbf{t^\ast}) & \Sigma_{d_m d_m^{\prime\prime}}(\mathbf{t^\ast},\mathbf{t^\ast})\\  \Sigma_{d_m d_m^\prime}(\mathbf{t^\ast},\mathbf{t^\ast})^T & \Sigma_{d_m^\prime}(\mathbf{t^\ast},\mathbf{t^\ast}) & \Sigma_{d_m^\prime d_m^{\prime\prime}}(\mathbf{t^\ast},\mathbf{t^\ast})\\ \Sigma_{d_m d_m^{\prime\prime}}(\mathbf{t^\ast}, \mathbf{t^\ast})^T & \Sigma_{d_m^\prime d_m^{\prime \prime}}(\mathbf{t^\ast}, \mathbf{t^\ast})^T & \Sigma_{d_m^{\prime\prime}}(\mathbf{t^\ast}, \mathbf{t^\ast})\end{bmatrix} 
\end{align*} Using properties of the multivariate normal distribution
the individual components are given by \begin{align*}
  \mu_{d_m}(\mathbf{t}^\ast) &= \mu_{\bm{\beta}_m}(\mathbf{t}^\ast) + C_{\bm{\theta}_m}(\mathbf{t}^\ast, \mathbf{t}_m)\left(C_{\bm{\theta}_m}(\mathbf{t}_m, \mathbf{t}_m) + \sigma^2_m I\right)^{-1}\left(\mathbf{D}_m - \mu_{\bm{\beta}_m}(\mathbf{t}_m)\right)\\
  \mu_{d_m^\prime}(\mathbf{t}^\ast) &= \mu^\prime_{\bm{\beta}_m}(\mathbf{t}^\ast) + \partial_1 C_{\bm{\theta}_m}(\mathbf{t}^\ast, \mathbf{t}_m)\left(C_{\bm{\theta}_m}(\mathbf{t}_m, \mathbf{t}_m) + \sigma^2_m I\right)^{-1}\left(\mathbf{D}_m - \mu_{\bm{\beta}_m}(\mathbf{t}_m)\right) \\
  \mu_{d_m^{\prime\prime}}(\mathbf{t}^\ast) &= \mu^{\prime\prime}_{\bm{\beta}_m}(\mathbf{t}^\ast) + \partial_1^2 C_{\bm{\theta}_m}(\mathbf{t}^\ast, \mathbf{t}_m)\left(C_{\bm{\theta}_m}(\mathbf{t}_m, \mathbf{t}_m) + \sigma^2_m I\right)^{-1}\left(\mathbf{D}_m - \mu_{\bm{\beta}_m}(\mathbf{t}_m)\right)\\
  \Sigma_{d_m}(\mathbf{t}^\ast, \mathbf{t}^\ast) &= C_{\bm{\theta}_m}(\mathbf{t}^\ast, \mathbf{t}^\ast) - C_{\bm{\theta}_m}(\mathbf{t}^\ast, \mathbf{t}_m)\left(C_{\bm{\theta}_m}(\mathbf{t}_m, \mathbf{t}_m) + \sigma^2_m I\right)^{-1} C_{\bm{\theta}_m}(\mathbf{t}_m, \mathbf{t}^\ast)\\
  \Sigma_{d_m^\prime}(\mathbf{t}^\ast, \mathbf{t}^\ast) &= \partial_1\partial_2C_{\bm{\theta}_m}(\mathbf{t}^\ast, \mathbf{t}^\ast) - \partial_1C_{\bm{\theta}_m}(\mathbf{t}^\ast, \mathbf{t}_m)\left(C_{\bm{\theta}_m}(\mathbf{t}_m, \mathbf{t}_m) + \sigma^2_m I\right)^{-1} \partial_2C_{\bm{\theta}_m}(\mathbf{t}_m, \mathbf{t}^\ast)\\
  \Sigma_{d_m^{\prime\prime}}(\mathbf{t}^\ast, \mathbf{t}^\ast) &= \partial_1^2\partial_2^2 C_{\bm{\theta}_m}(\mathbf{t}^\ast, \mathbf{t}^\ast) - \partial_1^2 C_{\bm{\theta}_m}(\mathbf{t}^\ast, \mathbf{t}_m)\left(C_{\bm{\theta}_m}(\mathbf{t}_m, \mathbf{t}_m) + \sigma^2_m I\right)^{-1} \partial_2^2 C_{\bm{\theta}_m}(\mathbf{t}_m, \mathbf{t}^\ast)\\
  \Sigma_{d_m d_m^\prime}(\mathbf{t}^\ast, \mathbf{t}^\ast) &= \partial_2 C_{\bm{\theta}_m}(\mathbf{t}^\ast, \mathbf{t}^\ast) - C_{\bm{\theta}_m}(\mathbf{t}^\ast, \mathbf{t}_m)\left(C_{\bm{\theta}_m}(\mathbf{t}_m, \mathbf{t}_m) + \sigma^2_m I\right)^{-1} \partial_2 C_{\bm{\theta}_m}(\mathbf{t}_m, \mathbf{t}^\ast)\\
  \Sigma_{d_m d_m^{\prime\prime}}(\mathbf{t}^\ast, \mathbf{t}^\ast) &= \partial_2^2 C_{\bm{\theta}_m}(\mathbf{t}^\ast, \mathbf{t}^\ast) - C_{\bm{\theta}_m}(\mathbf{t}^\ast, \mathbf{t}_m)\left(C_{\bm{\theta}_m}(\mathbf{t}_m, \mathbf{t}_m) + \sigma^2_m I\right)^{-1} \partial_2^2 C_{\bm{\theta}_m}(\mathbf{t}_m, \mathbf{t}^\ast)\\
  \Sigma_{d_m^\prime d_m^{\prime\prime}}(\mathbf{t}^\ast, \mathbf{t}^\ast) &= \partial_1 \partial_2^2 C_{\bm{\theta}_m}(\mathbf{t}^\ast, \mathbf{t}^\ast) - \partial_1 C_{\bm{\theta}}(\mathbf{t}^\ast, \mathbf{t}_m)\left(C_{\bm{\theta}_m}(\mathbf{t}_m, \mathbf{t}_m) + \sigma^2_m I\right)^{-1} \partial_2^2 C_{\bm{\theta}_m}(\mathbf{t}_m, \mathbf{t}^\ast)
\end{align*}

\hypertarget{nba-season-20192020-games-ordered-by-decreasing-median-posterior-excitement-trend-index}{%
\section{NBA season 2019--2020 games ordered by decreasing median
posterior Excitement Trend
Index}\label{nba-season-20192020-games-ordered-by-decreasing-median-posterior-excitement-trend-index}}

\begingroup\fontsize{8}{10}\selectfont

\begin{longtable}{rl|llr|rrr}
\toprule
No & Date & Home team & Away team & Median & Mean & 2.5\% & 97.5\%\\
\midrule
\endfirsthead
\multicolumn{8}{@{}l}{\textit{(continued)}}\\
\toprule
No & Date & Home team & Away team & Median & Mean & 2.5\% & 97.5\%\\
\midrule
\endhead

\endfoot
\bottomrule
\endlastfoot
1 & 2020-01-16 & New Orleans Pelicans & Utah Jazz & 26.23 & 26.36 & 19.28 & 33.44\\
2 & 2020-02-05 & Oklahoma City Thunder & Cleveland Cavaliers & 26.17 & 25.59 & 11.96 & 35.76\\
3 & 2020-03-04 & Cleveland Cavaliers & Boston Celtics & 26.11 & 26.27 & 21.90 & 31.00\\
4 & 2020-03-01 & Denver Nuggets & Toronto Raptors & 25.28 & 23.91 & 10.34 & 33.29\\
5 & 2019-11-05 & Oklahoma City Thunder & Orlando Magic & 24.93 & 23.53 & 7.40 & 31.60\\
6 & 2020-02-29 & San Antonio Spurs & Orlando Magic & 24.89 & 24.47 & 18.06 & 30.15\\
7 & 2020-02-12 & Memphis Grizzlies & Portland Trail Blazers & 24.34 & 23.97 & 17.99 & 29.33\\
8 & 2020-01-17 & Indiana Pacers & Minnesota Timberwolves & 23.76 & 23.89 & 14.57 & 33.28\\
\end{longtable}
\endgroup{}